\documentclass[aps,pr,onecolumn,
superscriptaddress,groupedaddress,nofootinbib,floatfix]{revtex4-2}  
\usepackage[colorlinks,citecolor=blue,urlcolor=blue,linkcolor=blue]{hyperref}
\usepackage{soul}
\usepackage{amsfonts}
\usepackage{amssymb}
\usepackage{amsmath}
\usepackage{amsthm}
\usepackage{graphicx}
\usepackage{appendix}
\usepackage{bbold}
\usepackage{dsfont}
\usepackage{enumitem}
\usepackage{xcolor}
\usepackage{yfonts}
\usepackage{MnSymbol}

\usepackage[normalem]{ulem}
\newcommand\redsout{\bgroup\markoverwith{\textcolor{red}{\rule[0.4ex]{3pt}{0.7pt}}}\ULon}

\providecommand{\U}[1]{\protect\rule{.1in}{.1in}}

\usepackage{color}
\usepackage{listings}
\definecolor{darkgreen}{rgb}{0,0.35,0}
\definecolor{Rood}{rgb}{1, 0, 0}

\begin{document}

\title{Fermionic quantum gas at finite temperature within a Lorentz violating background}

\author{Rafael L.~J.~Costa}
\email{rafaelljc@id.uff.br}
\affiliation{Instituto de F\'isica, Universidade Federal Fluminense, Campus da Praia Vermelha, Av. Litor\^anea s/n, 24210-346, Niter\'oi, RJ, Brazil}
\author{Rodrigo~F.~Sobreiro}
\email{rodrigo\_sobreiro@id.uff.br}
\affiliation{Instituto de F\'isica, Universidade Federal Fluminense, Campus da Praia Vermelha, Av. Litor\^anea s/n, 24210-346, Niter\'oi, RJ, Brazil}

\begin{abstract}
In this work we consider a fermionic quantum gas within a Lorentz-Violating background at finite temperature. We derive the effective action within Path Integral formalism considering the interaction of external electromagnetic field and Lorentz violating background fields with quantum fermions. To introduce the temperature effects, we employ the Matsubara formalism. Comments about the corresponding phenomenology are also made. 
\end{abstract}

\maketitle

\section{Introduction}

Lorentz symmetry (LS) is a fundamental principle in modern physics \cite{Bargmann:1946me,Bargmann:1948ck}, which states that the physical laws of physics should be the same for all observers who are moving at a constant velocity relative to each other. A direct breaking of LS would imply that this principle is not valid, and that different observers would see different physical laws. There are several ways in which LS could be broken \cite{Kostelecky:1989jp,Kostelecky:1988zi,Carroll:1989vb,Kostelecky:1989jw,Colladay:1996iz,Colladay:1998fq}. One possibility is through the existence of preferred directions in space or time, which would violate the isotropy and homogeneity of space and time, respectively. Another possibility is through the existence of preferred frame of reference, which would violate the principle of relativity. In recent years, there has been growing interest in studying the effects of Lorentz symmetry violation, which is a possible extension to the Standard Model of particle physics, see for instance
\cite{Jackiw:1999yp,Kostelecky:2003fs,Kostelecky:2005ic,Belich:2004ng,Mariz:2005jh,Gomes:2009ch,Diaz:2011ia,SSantos:2015mzs,Santos:2016bqc,Gomes:2019uay,Ferreira:2020wde,Filho:2021rin,Mariz:2021cik}. Lorentz violation can manifest in various ways, including modifications to the dispersion relations of particles \cite{Carroll:1989vb,Paixao:2023qvo}.

In this article, we will explore the connection between Lorentz-violating terms in the fermionic sector and the Fermi-Dirac distribution, examining the effects of Lorentz symmetry violation on the distribution of fermions with finite temperature. We will also discuss the potential implications of these findings for the development of new physics beyond the Standard Model.

To introduce the temperature effects, we will make use of the Matsubara formalism \cite{Matsubara:1955ws,Kapusta:2006pm}. Matsubara formalism is a powerful technique in statistical mechanics and quantum field theory for describing the thermodynamic properties of systems at finite temperature. Named after the physicist Tomonaga Matsubara, who introduced the method in 1955 \cite{Matsubara:1955ws}, Matsubara formalism provides a framework for computing thermal averages and response functions in many-body systems. The formalism involves a complexification of the time variable and a discrete Fourier transform, which allow one to relate the imaginary-time correlation functions of the system to its physical observables at finite temperature. This has many applications in condensed matter physics and high-energy physics, among other fields \cite{Litvinova:2021ygq,mena2020thermal,Hotta:2019dtk,Litvinova:2018nrg}.

Also, to derive the effects of temperature on a fermionic quantum gas with Lorentz-violating background, we consider the interaction of quantum Dirac fields with an external electromagnetic field so that the formalism of effective action can be used. In fact, we consider Dirac-Maxwell theory supplemented with Lorentz-violating terms at the fermionic sector. Eventually, in order to track down the effects of temperature and the background, the electromagnetic field will be driven to zero.

Our article is organized as follows. In section \ref{GT}, we present the general form of the effective action at finite temperature using regularization via Zeta function. In subsections \ref{PC}, \ref{PC2} and \ref{PC3}, we perform the specific calculation for the Lorentz-violation background fields $d^{\mu\nu}$, $b^{\mu}$, $c^{\mu\nu}$, and $f^{\mu}$, respectively. In section \ref{PHE}, we briefly comment on the phenomenology of the results. Finally, in section \ref{FINAL}, we present the conclusions of the work.

\section{One-loop Effective Lagrangian at Finite Temperature}\label{GT}

In the Path Integral formalism, the effective action can be obtained from the functional generator 
\begin{align}
Z\left[A,n\right]=\int\mathcal{D}\bar{\psi}\mathcal{D}\psi e^{i\int d^{4}x\mathcal{L}\left[A,n\right]}\;,\label{Z1}
\end{align}
with $\mathcal{L}\left[A,n\right]$ being the Lagrangian of the theory depending on the external electromagnetic potential\footnote{Greek indexes range from $0$ to $3$.} $A_\mu$ and the set of Lorentz-violating background fields $n$. The fields $\psi$ and $\bar{\psi}$ stand for Dirac spinors and their conjugate, respectively. The usual normalization factor in \eqref{Z1} is implicit.

We consider a QED extension which involves only Lorentz-violating terms in the fermionic sector. The general Lagrangian of this model is given by\footnote{Natural units, $\hbar=c=1$, are employed in the present work.} \cite{Colladay:1998fq,Kostelecky:2003fs}
\begin{align}
\label{L0}
    \mathcal{L}_{QED_{ex}}=-\frac{1}{4}\mathcal{F}_{\mu\nu}\mathcal{F}^{\mu\nu}+\bar{\psi}\left(\Gamma_{\mu}\Pi^{\mu}-M\right)\psi\, ,
\end{align}
where
\begin{eqnarray}
\label{LVfields}
    \Gamma_{\mu}&\equiv&\gamma_\mu+ c_{\nu\mu}\gamma^{\nu}+d_{\nu\mu}\gamma^{5}\gamma^{\nu}+e^{\mu}+if_{\mu}\gamma^{5}+\frac{1}{2}g_{\alpha\beta\mu}\sigma^{\alpha\beta}\;,\nonumber\\
    M&\equiv&m+ im_{5}\gamma_{5}+a^{\mu}\gamma_{\mu}+b^{\mu}\gamma_{5}\gamma_{\mu}+\frac{1}{2}h^{\mu\nu}\sigma_{\mu\nu}\;.
\end{eqnarray}
The parameters $m$ and $e$ stands for electron mass and charge, respectively. The covariant derivative is defined as $\Pi^{\mu}=i\partial^{\mu}-eA^{\mu}$, and the field strength is written as $\mathcal{F}_{\mu\nu}=\partial_{\mu}A_{\nu}-\partial_{\nu}A_{\mu}$. All tensor fields in \eqref{LVfields} are Lorentz-violating background fields belonging to the above defined set $n$. Clearly, they provide anisotropy to spacetime by defining privileged directions. The tensor fields in the first line of \eqref{LVfields} has no mass dimension while in the second line they have mass dimension $1$. Furthermore, tensor fields with an odd number of indexes do not preserve CPT while an even number of indexes preserve CPT, see for instance \cite{SSantos:2015mzs}. For the $\gamma$ matrices, we define them from the Clifford algebra \footnote{The Minkowski metric given by $\eta_{\mu\nu}=\mathrm{diag}\left( -1,1,1,1\right)$ and further conventions are the same as in \cite{Itzykson:1980rh}.} $\left\{ \gamma^{\mu},\gamma^{\nu}\right\}=-2\eta^{\mu\nu}$. Moreover,
\begin{eqnarray}
    \sigma^{\mu\nu}&=&\frac{i}{2}\left[\gamma^{\mu},\gamma^{\nu}\right]\,,\nonumber\\
    \gamma_{5}&=&-\frac{i}{4!}\epsilon^{\mu\nu\sigma\beta}\gamma_{\mu}\gamma_{\nu}\gamma_{\sigma}\gamma_{\beta}\,.
\end{eqnarray}

We remark once again that the fermionic field is the only quantum field in \eqref{L0}, all Lorentz-violating fields and also the gauge field $A_\mu$ are considered here as background fields.

Employing the usual Gaussian integral formula for the fermion field \cite{Itzykson:1980rh,Dittrich:1978fc}, we get
\begin{align}
\label{L1}
Z\left[A,n\right]&=\det\left(\Gamma^{\mu}\Pi_{\mu}-M\right)=\det G\left[A,n\right]\,
\end{align}
To evaluate the determinant in \eqref{L1}, we will use the method of regularization via the Zeta function \cite{Hawking:1976ja}. For an arbitrary operator $\Xi$, the determinant associated with it, can be calculated as
\begin{align}
    \label{L22}
    \det\Xi&=e^{-\partial_{\left(s\right)}\zeta_{\Xi}\left(0\right)}\,,
\end{align}
where $\zeta_{\Xi}\left(s\right)$ is the Zeta function of the operator $\Xi$. The advantage of using this determinant definition is that $\partial_{\left(s\right)}\zeta_{\Xi}\left(0\right)$ is not singular for most operators of physical interest.

Therefore, writing the one-loop effective action, $iW\left[A\right]^{\left(1\right)}$, we have
\begin{align}
\label{L3}
iW\left[A\right]^{\left(1\right)}=\ln\det G\left[A,n\right]\,.
\end{align}
Taking advantage from the fact that the determinant of an operator $\Xi$ is invariant under a similarity transformation $S$, as follows, $\det \left[\Xi\right]=\det \left[S\Xi S^{-1}\right]=\det \left[\Xi S\Xi S^{-1}\right]^{\frac{1}{2}}$, we can rewrite \eqref{L3} as
\begin{align}
    \label{L4}
iW\left[A\right]^{\left(1\right)}=\frac{1}{2}\ln\det\left\{ G\left[A,n\right]G^{\left(S\right)}\left[A,n\right]\right\}  \,,
\end{align}
where $G^{\left(S\right)}\left[A,n\right]=SG\left[A,n\right]S^{-1}$. This will be useful for calculating the Zeta functions of the operators studied here. In the next sections we will work some particular cases of \eqref{LVfields}.

\subsection{The $d^{\mu\nu}$ contribution}\label{PC}

We start our investigation by only looking at the contribution of the $d^{\mu\nu}$ field. The effective action is evaluated by \eqref{L4}, where the charge conjugation $C$ matrix is chosen as the similarity transformation. Thence, for the $d^{\mu\nu}$, the effective action is written as 
\begin{align}
    \label{L5}
iW\left[A,d\right]^{\left(1\right)}&=\ln\det\left\{ G\left[A,d\right]G^{\left(C\right)}\left[A,d\right]\right\} ^{\frac{1}{2}}\,\,,
\end{align}
with $G\left[A,d\right]=\Gamma^{\mu}\Pi_{\mu}-m$, and
\begin{align}
G\left[A,d\right]&=\Gamma^{\mu}\Pi_{\mu}-m\,,\nonumber\\
\Gamma^{\mu}&=\gamma^{\mu}+d^{\beta\mu}\gamma^{5}\gamma_{\beta}\,,\nonumber\\
M&=m\,.
\end{align}
In particular, for the sake of simplicity, we choose $d^{\mu\nu}$ to be time-like by setting $d^{\mu\nu}=\textrm{diag}(d,0,0,0)$. And, using $\ln\det\Xi=Tr\ln\Xi$, one achieves
\begin{align}
    \label{L6}
iW\left[A,d\right]^{\left(1\right)}&=\frac{1}{2}Tr\ln\left[\left(\Gamma^{\mu}\Pi_{\mu}-m\right)\left(-\Gamma^{\mu}\Pi_{\mu}-m\right)\right]\,.
\end{align}
Now, developing \eqref{L6} a bit more, we arrive at

\begin{eqnarray}
    \label{L66}
iW\left[A,d\right]^{\left(1\right)}&=\frac{1}{2}Tr\ln\left[-\Pi^{2}+\frac{e}{2}\sigma_{\mu\xi}\mathcal{F}^{\mu\xi}+id\gamma^{5}\sigma_{0}^{\,1}\Pi_{0}\Pi_{1}+id\gamma^{5}\sigma_{0}^{\,2}\Pi_{0}\Pi_{2}+id\gamma^{5}\sigma_{0}^{\,3}\Pi_{0}\Pi_{3}+\right.\nonumber\\&\left.+id\gamma^{5}\sigma_{0}^{\,1}\Pi_{1}\Pi_{0}+id\gamma^{5}\sigma_{0}^{\,2}\Pi_{2}\Pi_{0}+id\gamma^{5}\sigma_{0}^{\,3}\Pi_{3}\Pi_{0}-d^{2}\Pi_{0}\Pi_{0}+m^{2}\right]\,.
\end{eqnarray}

At this point, it is convenient to take the limit of vanishing electromagnetic field, imposing $\mathcal{F}^{\mu\xi}=0$. Thus,
\begin{align}
    \label{L7}
iW\left[A,d\right]^{\left(1\right)}&=\frac{1}{2}Tr\ln K,
\end{align}
where 
\begin{align}
\label{K}
    K&=\begin{pmatrix}\alpha & i\beta+\gamma & i\kappa & 0\\
i\beta-\gamma & \alpha & 0 & -i\kappa\\
i\kappa & 0 & \alpha & i\beta+\gamma\\
0 & -i\kappa & i\beta-\gamma & \alpha
\end{pmatrix}\,,\nonumber\\\alpha&=\partial_{0}^{2}-\partial_{\left(3\right)}^{2}-d^{2}\partial_{0}^{2}+m^{2}\,,\nonumber\\\beta&=2d\partial_{1}\partial_{0}\,,\nonumber\\\gamma&=2d\partial_{2}\partial_{0}\,,\nonumber\\\kappa&=2d\partial_{3}\partial_{0}\,.
\end{align}
Diagonalizing $K$ and moving to momentum space, $\partial_{\mu}\to-ik_{\mu}$, $K$ is rewritten as
\begin{align}
\label{K2}
K_{\left(diag\right)}&=\begin{pmatrix}\alpha-\mathcal{P}^{*} & 0 & 0 & 0\\
0 & \alpha-\mathcal{P}^{*} & 0 & 0\\
0 & 0 & \alpha+\mathcal{P}^{*} & 0\\
0 & 0 & 0 & \alpha+\mathcal{P}^{*}
\end{pmatrix}\,,
\end{align}
with $\mathcal{P}^{*}=-2dk_{0}k$. Thence, expression \eqref{L7} yields
\begin{align}
    \label{L8}
iW\left[A,d\right]^{\left(1\right)}&=tr_{x}\ln\left[\mu^{-2}\left(\alpha-\mathcal{P}^{*}\right)\right]+tr_{x}\ln\left[\mu^{-2}\left(\alpha+\mathcal{P}^{*}\right)\right]\,,
\end{align}
where $tr_{x}$ indicates that the trace is in the coordinates space. The arbitrary mass parameter $\mu$ was introduced in order to keep the action dimensionless, by definition. Now, using definitions \eqref{L1} and \eqref{L22}, one gets
\begin{align}
\label{L9}
    iW\left[A,d\right]^{\left(1\right)}&=-\partial_{\left(s\right)}\zeta_{2}\left(0\right)\,,
\end{align}
where $\zeta_{2}\left(s\right)=\zeta_{\mu^{-2}\left(\alpha-\mathcal{P}^{*}\right)}+\zeta_{\mu^{-2}\left(\alpha+\mathcal{P}^{*}\right)}$. If we assume that the fields are normalized on a volume $\Omega=L^{4}$ in 4-dimensional Euclidean space, making a Wick rotation $\partial_{0}\to i\partial_{0}$, we can approximate $\left(-\partial_{\left(t\right)}^{2} -\partial_{\left(3\right)}^{2}\right)$ by plane waves with eigenvalues $\left(k_{0}^{2}+k_{3}^{2}\right)$. Hence, we can finally write our function $\zeta_{2}\left(s\right)$,
\begin{align}
    \label{L10}
\zeta_{2}\left(s\right)&=\mu^{2s}\Omega\left[I_{+}+I_{-}\right]\,,
\end{align}
where 
\begin{align}
    \label{int1}
  I_{+}&=\int\frac{d^4k}{\left(2\pi\right)^{3}}\left[\left(k_{0}^{2}+k^{2}+d^{2}k_{0}^{2}+m^{2}+2dk_{0}k\right)^{-s}\right]\nonumber\\I_{-}&=\int\frac{d^4k}{\left(2\pi\right)^{3}}\left[\left(k_{0}^{2}+k^{2}+d^{2}k_{0}^{2}+m^{2}-2dk_{0}k\right)^{-s}\right]\,,
\end{align}
with $k=k_{\left(3\right)}$ being the three dimensional momentum.

To account for temperature effects in the function $\zeta_{2,}\left(s\right)$ for a fermionic quantum gas we replace the function of $k_{0}$ by an anti-periodic function of temperature, known as finite-temperature Matsubara prescription \cite{Kapusta:2006pm},
\begin{align}
    \label{matsu}
    \int_{-\infty}^{\infty}\frac{dk_{0}}{2\pi}f\left(k_{0}^{2}\right)&\rightarrow\sum_{l=0}^{\infty}f\left(\frac{\pi^{2}}{\beta^{2}}\left[2l+1\right]^{2}\right)\,,
\end{align}
with $\beta=\frac{1}{k_{b}T}$, $k_{b}$ being the Boltzman constant, and $T$ standing for temperature. So, the integrals \eqref{int1} become

\begin{eqnarray}
    \label{int2}
I_{+}&=\sum_{l=0}^{\infty}\int_{0}^{2\pi}\int_{0}^{\pi}\int_{0}^{\infty}\frac{k^{2}\sin\theta dkd\theta d\phi}{\left(2\pi\right)^{3}}\left(k_{0}^{2}+k^{2}+d^{2}k_{0}^{2}+m^{2}+2dk_{0}k\right)^{-s}\,,\nonumber\\I_{-}&=\sum_{l=0}^{\infty}\int_{0}^{2\pi}\int_{0}^{\pi}\int_{0}^{\infty}\frac{k^{2}\sin\theta dkd\theta d\phi}{\left(2\pi\right)^{3}}\left(k_{0}^{2}+k^{2}+d^{2}k_{0}^{2}+m^{2}-2dk_{0}k\right)^{-s}\,,
\end{eqnarray}

where spherical coordinates, $dk^{3}=k^{2}\sin\theta dkd\theta d\phi$, were used.

Let us working first on the computation of $I_{+}$. We can rewrite it as
\begin{eqnarray}
    \label{int3} I_{+}&=\sum_{l=0}^{\infty}\int_{0}^{2\pi}\int_{0}^{\pi}\int_{0}^{\infty}\frac{k^{2}\sin\theta dkd\theta d\phi}{\left(2\pi\right)^{3}}\times\nonumber\\&\times\left\{ \Delta+\left[\sigma\left(l+\frac{1}{2}\right)d+k\right]^{2}\right\} ^{-s}\,,
\end{eqnarray}
with $\sigma=\frac{2\pi}{\beta}$ and $\Delta=\sigma^{2}\left(l+\frac{1}{2}\right)^{2}+m^{2}$. So, performing the $\theta$ and $\phi$ integration, we arrive at
\begin{align}
        \label{int4} 
  I_{+}&=\frac{4\pi}{\left(2\pi\right)^{3}}\sum_{l=0}^{\infty}\Delta^{-s}\int_{0}^{\infty}k^{2}dk\left\{ 1+\left[\frac{\sigma\left(l+\frac{1}{2}\right)d+k}{\Delta^{\frac{1}{2}}}\right]^{2}\right\} ^{-s}\,.
\end{align}
Now, making the following change of variables
\begin{align}
x&=\frac{k+\sigma\left(l+\frac{1}{2}\right)d}{\Delta^{\frac{1}{2}}}\,,
\end{align}
such that, \eqref{int4} becomes
\begin{align}
    \label{int5}
    I_{+}&=\frac{4\pi}{\left(2\pi\right)^{3}}\sum_{l=0}^{\infty}\Delta^{\frac{3}{2}-s}\int_{0}^{\infty}\left(x-\kappa_{l}\right)^{2}\left(1+x^{2}\right)^{-s}dx\,,
\end{align}
where $\kappa_{l}=\frac{\sigma\left(l+\frac{1}{2}\right)d}{\Delta^{\frac{1}{2}}}$, the integral \eqref{int5} can be solved with the help of Mathematica \cite{Mathematica} and the result is
\begin{eqnarray}
    \label{int6}
 I_{+}&=\frac{4\pi}{\left(2\pi\right)^{3}}\sum_{l=0}^{\infty}\left\{ \frac{\Delta^{1-s}\sigma\left(l+\frac{1}{2}\right)d}{\left(1-s\right)}+\right.\left.\frac{\pi^{\frac{1}{2}}\left\{ \Delta^{\frac{3}{2}-s}+\left(-3+2s\right)\Delta^{\frac{1}{2}-s}\sigma^{2}\left(l+\frac{1}{2}\right)^{2}d^{2}\right\} \Gamma\left(-\frac{3}{2}+s\right)}{4\Gamma\left(s\right)}\right\} \,,
\end{eqnarray}
with $\Gamma\left(s\right)$ being the Gamma function. Now, since $d$ is small, we can expand \eqref{int6} to obtain
\begin{eqnarray}
\label{int7}
I_{+}&=\frac{4\pi}{\left(2\pi\right)^{3}}\sum_{l=0}^{\infty}\left\{ \frac{\Delta^{\frac{3}{2}-s}\Gamma\left(\frac{3}{2}\right)\Gamma\left(-\frac{3}{2}+s\right)}{2\Gamma\left(s\right)}+\frac{\Delta^{1-s}\sigma\left(l+\frac{1}{2}\right)d}{\left(1-s\right)}\right.+\left.\frac{\pi^{\frac{1}{2}}\left(-3+2s\right)\Delta^{\frac{1}{2}-s}\sigma^{2}\left(l+\frac{1}{2}\right)^{2}d^{2}\Gamma\left(-\frac{3}{2}+s\right)}{4\Gamma\left(s\right)}+\mathcal{O}\left(d^{3}\right)\right\}\,.
\end{eqnarray}
At this point, it is convenient to take the zero mass limit in order to obtain a closed expression for the sums in \eqref{int7}. Thus, 
\begin{eqnarray}
\label{int8}
I_{+}&=\frac{4\pi}{\left(2\pi\right)^{3}}\sum_{l=0}^{\infty}\left\{ \frac{\left[\sigma^{2}\left(l+\frac{1}{2}\right)^{2}\right]^{\frac{3}{2}-s}\Gamma\left(\frac{3}{2}\right)\Gamma\left(-\frac{3}{2}+s\right)}{2\Gamma\left(s\right)}+\right.\frac{\left[\sigma^{2}\left(l+\frac{1}{2}\right)^{2}\right]^{1-s}\sigma\left(l+\frac{1}{2}\right)d}{\left(1-s\right)}+\left.\frac{\left(-3+2s\right)\left[\sigma^{2}\left(l+\frac{1}{2}\right)^{2}\right]^{\frac{1}{2}-s}\sigma^{2}\left(l+\frac{1}{2}\right)^{2}d^{2}\Gamma\left(\frac{3}{2}\right)\Gamma\left(-\frac{3}{2}+s\right)}{2\Gamma\left(s\right)}\right\}\,.
\end{eqnarray}

For $I_{-}$, the procedure is the same, changing only the sign in the term of $\kappa_{l}$, in such a way that the result is 
\begin{eqnarray}
\label{int9}
I_{-}&=\frac{4\pi}{\left(2\pi\right)^{3}}\sum_{l=0}^{\infty}\left\{ \frac{\left[\sigma^{2}\left(l+\frac{1}{2}\right)^{2}\right]^{\frac{3}{2}-s}\Gamma\left(\frac{3}{2}\right)\Gamma\left(-\frac{3}{2}+s\right)}{2\Gamma\left(s\right)}\right.-\frac{\left[\sigma^{2}\left(l+\frac{1}{2}\right)^{2}\right]^{1-s}\sigma\left(l+\frac{1}{2}\right)d}{\left(1-s\right)}+\left.\frac{\left(-3+2s\right)\left[\sigma^{2}\left(l+\frac{1}{2}\right)^{2}\right]^{\frac{1}{2}-s}\sigma^{2}\left(l+\frac{1}{2}\right)^{2}d^{2}\Gamma\left(\frac{3}{2}\right)\Gamma\left(-\frac{3}{2}+s\right)}{2\Gamma\left(s\right)}\right\} \,.
\end{eqnarray}
We notice that the only difference between \eqref{int8} and \eqref{int9} is a minus sign in the first order terms in $d$. Therefore, the lowest contribution in $d$ is of second-order. From the two-argument Zeta function definition
\begin{align}
    \label{zeta}
\zeta\left(z,p\right)&=\sum_{n=0}^{\infty}\left(n+p\right)^{-z}\,,
\end{align}
we can write the final form of \eqref{L10},
\begin{eqnarray}
\label{L11}
\zeta_{2}\left(s\right)&=\mu^{2s}\Omega\frac{4\pi}{\left(2\pi\right)^{3}}\frac{\Gamma\left(\frac{3}{2}\right)\Gamma\left(-\frac{3}{2}+s\right)\sigma^{3-2s}\zeta\left(2s-3,\frac{1}{2}\right)}{\Gamma\left(s\right)}\left[1+\left(-3+2s\right)d^{2}\right]\,.
\end{eqnarray}

Now we finally proceed to computed the effective Lagrangian. Its relation with the effective action is given by
\begin{align}
iW\left[A\right]^{\left(1\right)}&=\int dx^{4}\mathcal{L}\left(\beta,d\right)^{\left(1\right)}=\Omega_{\left(3\right)}\beta\mathcal{L}\left(\beta,d\right)^{\left(1\right)}\,.
\end{align}
Comparing this expression with expression \eqref{L9} we infer
\begin{align}
    \label{L12}
\mathcal{L}^{\left(1\right)}\left(\beta,d\right)&=-\Omega_{\left(3\right)}^{-1}\beta^{-1}\partial_{\left(s\right)}\zeta_{2}^{\beta}\left(0\right)\,.
\end{align}
Thus, performing the derivative on $s$ and then taking the limit $s\to0$, we get
\begin{align}
    \label{L13}
\mathcal{L}^{\left(1\right)}\left(\beta,d\right)&=-\beta^{-1}\frac{4\pi}{\left(2\pi\right)^{3}}\Gamma\left(\frac{3}{2}\right)\Gamma\left(-\frac{3}{2}\right)\sigma^{3}\zeta\left(-3,\frac{1}{2}\right)\left(1-3d^{2}\right)\,.
\end{align}
From
\begin{align}
\label{def2}
    \zeta\left(-n,q\right)&=-\frac{\partial_{\left(q\right)}B_{n+2}\left(q\right)}{\left(n+1\right)\left(n+2\right)}\,,
\end{align}
where $B_{n+2}\left(q\right)$ are the Bernoulli polynomials, which the first ones are
\begin{align}
\label{b1}
B_{0}\left(q\right)&=1\,,\nonumber\\B_{1}\left(q\right)&=q-\frac{1}{2}\,,\nonumber\\B_{2}\left(q\right)&=q^{2}-q+\frac{1}{6}\,,\nonumber\\B_{3}\left(q\right)&=q^{3}-\frac{3}{2}q^{2}+\frac{1}{2}q\,,\nonumber\\B_{4}\left(q\right)&=q^{4}-2q^{3}+q^{2}-\frac{1}{30}\,,\nonumber\\B_{5}\left(q\right)&=q^{5}-\frac{5}{2}q^{4}+\frac{5}{3}q^{3}-\frac{1}{6}q\,,
\end{align}
we have
\begin{align}
    \label{L14}
\mathcal{L}^{\left(1\right)}\left(\beta,d\right)&=\pi^{2}\frac{8}{3}\left.\frac{\partial_{\left(q\right)}B_{5}\left(q\right)}{20}\right|_{q=\frac{1}{2}}\left(\frac{1}{\beta}\right)^{4}\left(1-3d^{2}\right)\,.
\end{align}
Moreover, employing the useful relation of the derivative of the Bernoulli polynomials
\begin{align}
\label{b2}
\partial_{\left(q\right)}B_{n}\left(q\right)&=nB_{n-1}\left(q\right)\,,
\end{align}
we finally get\footnote{Here we have multiplied the result by a factor $2$ in order to account for the two spin projections of electrons and positrons gas.}
\begin{align}
    \label{L15}    \mathcal{L}^{\left(1\right)}\left(T,d\right)&=\frac{7}{180}\pi^{2}k_{b}^{4}T^{4}\left(1-3d^{2}\right)\,,
\end{align}

Expression \eqref{L15} can also be rewritten in the form\footnote{The well known identity
\begin{align}
    \label{def3}
\int_{0}^{\infty}\frac{\epsilon^{2n-1}d\epsilon}{e^{p\epsilon}+1}&=\left(1-2^{1-2n}\right)\left(\frac{2\pi}{p}\right)^{2n}\frac{\left|B_{2n}\right|}{4n}\,,
\end{align}
was used.}
\begin{align}
    \label{L16}
\mathcal{L}^{\left(1\right)}\left(T,d\right)&=\frac{2}{3\pi^{2}}\left(1-3d^{2}\right)\int_{0}^{\infty}\frac{\epsilon^{3}d\epsilon}{e^{\beta\epsilon}+1}\,,
\end{align}
which is the Fermi-Dirac distribution corrected by the Lorentz-violating background term.

\subsection{The $b^{\mu}$ contribution}\label{PC2}

As a second example, we study the $b^{\mu}$ contribution to the fermionic quantum gas. Specifically, for the sake of simplicity, the time-like contribution, $b^{\mu}=\left(b,0,0,0\right)$. The structure of calculation is the same as the previous one previous. Therefore, we will be quicker in calculations here. Starting with the effective action, it is given by
\begin{eqnarray}
    \label{LL1}    iW\left[A,b\right]^{\left(1\right)}&=\frac{1}{2}Tr\ln\left(\Pi^{2}-2ib\gamma_{5}\sigma_{01}\Pi_{1}-2ib\gamma_{5}\sigma_{02}\Pi_{2}\right.\left.-2ib\gamma_{5}\sigma_{03}\Pi_{3}-e\sigma^{3}\mathcal{B}_{z}+b^{2}+m^{2}\right)\,.
\end{eqnarray}
Again, the construction process of the Zeta function is analogous to the one discussed for the field $d^{\mu\nu}$. The limit of vanishing electromagnetic field is taken and we also move to the 4-dimensional Euclidean space. The Zeta function in this case reads
\begin{align}
\label{LL2}
\zeta_{2}^{\left(b\right)}\left(s\right)&=\mu^{2s}\Omega\left(I_{-}+I_{+}\right)\,,
\end{align}
with
\begin{eqnarray}
\label{LL3}
I_{-}^{\left(b\right)}&=\int\frac{d^4k}{\left(2\pi\right)^{4}}\left(m^{2}+b^{2}+k_{0}^{2}+k_{1}^{2}+\right.\left.+k_{2}^{2}+k_{3}^{2}-2b\sqrt{k_{1}^{2}+k_{2}^{2}+k_{3}^{2}}\right)^{-s}\,,\nonumber\\I_{+}^{\left(b\right)}&=\int\frac{d^4k}{\left(2\pi\right)^{4}}\left(m^{2}+b^{2}+k_{0}^{2}+k_{1}^{2}+\right.\left.+k_{2}^{2}+k_{3}^{2}+2b\sqrt{k_{1}^{2}+k_{2}^{2}+k_{3}^{2}}\right)\,.
\end{eqnarray}
Employing the Matsubara prescription \eqref{matsu} again, we have for $I_-^{(b)}$,
\begin{align}
\label{LL4}
I_{-}^{\left(b\right)}&=\frac{4\pi}{\left(2\pi\right)^{3}}\sum_{l=0}^{\infty}\int_{0}^{\infty}dkk^{2}\left[\Delta+\left(k-b\right)^{2}\right]^{-s}\,.
\end{align}
To solve \eqref{LL4}, the substitution $x=\frac{\left(k-b\right)}{\Delta_{-}^{\frac{1}{2}}}$ is performed, so \eqref{LL4} reads
\begin{align}
\label{LL4a}
I_{-}^{\left(b\right)}&=\frac{4\pi}{\left(2\pi\right)^{3}}\sum_{l=0}^{\infty}\Delta_{-}^{\frac{3}{2}-s}\int_{0}^{\infty}dx\left(x+b\right)^{2}\left(1+x^{2}\right)^{-s}\,.
\end{align}
Which can be solve, once again, with the help of Mathematica \cite{Mathematica}. The result is thus
\begin{eqnarray}
\label{LL5}
 I_{-}^{\left(b\right)}&=\frac{4\pi}{\left(2\pi\right)^{3}}\sum_{l=0}^{\infty}\Delta^{\frac{3}{2}-s}\left\{ \frac{b}{s-1}+\frac{\pi^{\frac{1}{2}}\left[1+b^{2}\left(2s-3\right)\right]\Gamma\left(s-\frac{3}{2}\right)}{4\Gamma\left(s\right)}\right\}\,\,\,.
\end{eqnarray}
Now, taking the limit of vanishing mass and using \eqref{zeta}, we have
\begin{eqnarray}
\label{LL6a}
I_{-}^{\left(b\right)}&=\frac{4\pi}{\left(2\pi\right)^{3}}\left(\frac{2\pi}{\beta}\right)^{3-2s}\zeta\left(2s-3,\frac{1}{2}\right)\left\{ \frac{b}{s-1}+\frac{\pi^{\frac{1}{2}}\left[1+b^{2}\left(2s-3\right)\right]\Gamma\left(s-\frac{3}{2}\right)}{4\Gamma\left(s\right)}\right\}\,.
\end{eqnarray}

For $I_{+}^{\left(b\right)}$ the process is analogous and the result is 
\begin{eqnarray}
\label{LL6}
I_{+}^{\left(b\right)}&=\frac{4\pi}{\left(2\pi\right)^{3}}\left(\frac{2\pi}{\beta}\right)^{3-2s}\zeta\left(2s-3,\frac{1}{2}\right)\left\{ -\frac{b}{s-1}+\frac{\pi^{\frac{1}{2}}\left[1+b^{2}\left(2s-3\right)\right]\Gamma\left(s-\frac{3}{2}\right)}{4\Gamma\left(s\right)}\right\}\,.
\end{eqnarray}

Combining these results in \eqref{LL2}, performing the $s$ derivative and taking $s\to0$, we arrive at
\begin{align}
    \label{LL7}
\partial_{\left(s\right)}\zeta_{2}^{\beta}\left(0\right)&=4\pi\Omega\left(\frac{1}{\beta}\right)^{3}\Gamma\left(\frac{3}{2}\right)\Gamma\left(-\frac{3}{2}\right)\zeta\left(-3,\frac{1}{2}\right)\left(1+b^{2}\right)\,.
\end{align}
Finally, using the relations \eqref{def2}, \eqref{b1} and \eqref{b2} together with \eqref{L12}, we get
\begin{align}
     \label{LL8}
\mathcal{L}^{\left(1\right)}\left(\beta,b\right)&=\frac{7}{180}\pi^{2}k_{b}^{4}T^{4}\left(1+b^{2}\right)\,,
\end{align}
where again we multiplied by a factor $2$ to account for the two spin projections of electrons and positrons. Making use of the identity \eqref{def3}, expression \eqref{LL8} can be rewritten as
\begin{align}
     \label{LL9}
\mathcal{L}^{\left(1\right)}\left(\beta,b\right)&=\frac{2}{3\pi^{2}}\left(1+b^{2}\right)\int_{0}^{\infty}\frac{\epsilon^{3}d\epsilon}{e^{\beta\epsilon}+1}\,.
\end{align}

We can compare \eqref{L16} and \eqref{LL9} and notice that both fields $b$ and $d$ generate the same physical effects for the Fermi-Dirac distribution. We can see a competition between then, being the first positive and the last negative. Perhaps, this competition may null the Lorentz-violating effect or almost null, depending of the values of the fields $d$ and $b$. Some Lorentz-Violating fields can represent the same physical effect and is possible to transit between then making a redefinition in the fermion fields, as we will see in the next section the $c^{\mu\nu}$ and $f^{\mu}$ has this property \cite{Altschul:2006ts,Karki:2022hzm}. Here we are treating the $d^{\mu\nu}$ and $b^{\mu}$ contribution as a competition, besides they come up with the same physical effect, because the analysis if is possible by a fermion field redefinition, to change the theory containing the one of them to another, is not done yet.

\subsection{The $c^{\mu\nu}$ and $f^{\mu}$ contribution}\label{PC3}

As a final example we consider the $c^{\mu\nu}$ and $f^{\mu}$ contributions for the fermionic quantum gas. These background fields have an interesting relationship. It is possible, from a transformation of the fermionic fields, to pass from a theory containing $f^{\mu}$ to a theory containing only $c^{\mu\nu}$, see \cite{Altschul:2006ts,Karki:2022hzm}.It is therefore expected that both fields generate the same physical effects. In fact, the relation between these fields is found to be
\begin{align}
    \label{LLL6}
    c^{\mu\nu}&=-\frac{1}{2}f^{\mu}f^{\nu}\,.
\end{align}
In the present discussion, we consider only time-like fields of the form, $c^{\mu\nu}=\textrm{diag}(c,0,0,0)$ and $f^{\mu}=(f,0,0,0)$. Thus, 
\begin{align}
    \label{LLL6a}
    c&=-\frac{1}{2}f^2\,.
\end{align}

Let us begin with $f^{\mu}$. The effective action for this case reads
\begin{align}
\label{LLL1}
iW\left[A,f\right]^{\left(1\right)}&=\ln\det G\left[A,f\right]\,,
\end{align}
where $G\left[A,f\right]=\left( -\partial^{2}+f^{2}\partial_{0}^{2}+m^{2}\right)^{\frac{1}{2}}$.
Passing to momentum space and evaluating the trace, the Zeta function is written as 
\begin{align}
    \label{LLL2}
\zeta_{2}\left(s\right)&=\frac{8\pi}{\left(2\pi\right)^{3}}\mu^{2s}\Omega\sum_{l=0}^{\infty}\int_{0}^{\infty}k^{2}dk\left(\Delta+k^{2}\right)^{-s}\,,
\end{align}
with $\Delta^{\left(f\right)}=\left(1-f^{2}\right)\sigma^{2}\left(l+\frac{1}{2}\right)^{2}+m^{2}$. Substituting $x=\frac{k}{\Delta^{\frac{1}{2}}}$ in expression \eqref{LLL2}, one gets
\begin{align}
\label{LLL2a}
\zeta_{2}\left(s\right)&=\frac{8\pi}{\left(2\pi\right)^{3}}\mu^{2s}\Omega\sum_{l=0}^{\infty}\Delta^{\frac{3}{2}-s}\int_{0}^{\infty}x^{2}\left(1+x^{2}\right)^{-s}dx\,.
\end{align}
The solution of the integral above reads \cite{Gradshteyn:1943cpj}
\begin{align}
    \label{deff1}
    \int_{0}^{\infty}dxx^{\mu-1}\left(1+x^{2}\right)^{\nu-1}&=\frac{1}{2}B\left(\frac{\mu}{2},1-\nu-\frac{\mu}{2}\right)\,,
\end{align}
where $B\left(\frac{\mu}{2},1-\nu-\frac{\mu}{2}\right)$ is Euler's beta function, also represented by
\begin{align}
    \label{deff2}
B\left(x,y\right)&=\frac{\Gamma\left(x\right)\Gamma\left(y\right)}{\Gamma\left(x+y\right)}\,.
\end{align}

Taking again the limit $m=0$ to obtain a closed form for the sum, expression \eqref{LLL2a} simplifies to
\begin{align}
\label{LLL3}
\zeta_{2}\left(s\right)&=\frac{4\pi}{\left(2\pi\right)^{3}}\mu^{2s}\Omega\left[\left(1-f^{2}\right)\sigma^{2}\right]^{\frac{3}{2}-s}\zeta\left(2s-3,\frac{1}{2}\right)\frac{\Gamma\left(\frac{3}{2}\right)\Gamma\left(s-\frac{3}{2}\right)}{\Gamma\left(s\right)}\,.
\end{align}
Using the relations \eqref{def2}, \eqref{b1} and \eqref{b2} together with \eqref{L12}, applying the $s$ derivative and taking the limit $s\to0$, we finally get
\begin{align}
    \label{LLL4}
\mathcal{L}^{\left(1\right)}\left(\beta,f\right)&=\frac{2}{3\pi^{2}}\left(1-f^{2}\right)^{\frac{3}{2}}\int_{0}^{\infty}\frac{\epsilon^{3}d\epsilon}{e^{\beta\epsilon}+1}\,.
\end{align}

For $c$ field, the calculation is basically the same and simply omit it. The final result is just
\begin{align}
    \label{LLL5}
\mathcal{L}^{\left(1\right)}\left(\beta,c\right)&=\frac{2}{3\pi^{2}}\left(1+2c+c^{2}\right)^{\frac{3}{2}}\int_{0}^{\infty}\frac{\epsilon^{3}d\epsilon}{e^{\beta\epsilon}+1}\,,
\end{align}
which agrees with \eqref{LLL4} at first order in $c$, in acordance with \eqref{LLL6} and \eqref{LLL6a}.

\section{Phenomenology}\label{PHE}

Let us discuss qualitatively a little more about the phenomenology of the results obtained above. First, we can note that both Lorentz-Violating fields add a very small contribution, as they should, as the fields themselves are very small \cite{kostelecky2011data}. But we can draw other insights from the results obtained. Also, as previously commented, we can notice that the fields $b^{\mu}$ and $d^{\mu\nu}$ produce similar contributions, in such a way that the physical effects are similar, one being positive and the other negative though. Having an competition between the fields when considered together.

In the case of fields $f^{\mu}$ and $c^{\mu\nu}$, this behavior was already expected due to the equivalence between them shown in \cite{Altschul:2006ts,Karki:2022hzm}. We can see then that these fields act in a similar way, in the thermal behavior of a fermionic quantum gas, at least for their time components.

Finally, we see that despite the Lorentz-Violating fields presenting a behavior similar to each other in the Fermi-Dirac distribution, they are very small, making the experimental analysis difficult. Even in different temperature regimes, we cannot draw significant effects because, looking at the results \eqref{L16}, \eqref{LL9}, \eqref{LLL4} and \eqref{LLL5}, we have that both Lorentz-Violating fields are $\ll1$, making the effects not so noticeable.

Furthermore, besides the fact that $b^{\mu}$ and $d^{\mu\nu}$ have the same shape, we cannot affirm that is possible to make a redefinition in the fermion fields and change the theory containing one of them to another. Here we only can notice that the time components of $b^{\mu}$ and $d^{\mu\nu}$ contributes to Fermi-Dirac distribution in a similar way, with one being positive contribution and the other a negative contribution, making in the first look a competition between them.

\section{Conclusions}\label{FINAL}

In this article, we have analyzed how the Lorentz violation modifies the Fermi-Dirac thermal distribution. For this, we use the model with violation only in the fermionic sector and employ the effective action formalism to calculate the effective Lagrangian density and the Matsubara formalism to include finite temperature effects. We also use the electromagnetic and Lorentz-violating fields as background fields. However, during the calculations, the vanishing of the electromagnetic field was used so that only the effects of temperature and Lorentz violation remains. Specifically, we calculated the effects for the time components of the fields $b^{\mu}$, $d^{\mu\nu}$, $f^{\mu}$ and $c^{\mu\nu}$.

The contribution of Lorentz-violating background fields for the Fermi-Dirac distribution is displayed in \eqref{L16}, \eqref{LL9}, \eqref{LLL4} and \eqref{LLL5}. As previously commented, this contributions are very small, since the fields themselves are small.

We can see from \eqref{L16} and \eqref{LL9} that the effects are manifested in the same shape for the $b^{\mu}$ and $d^{\mu\nu}$, being one positive and the other negative. This property represents a certain competition between the fields $b^{\mu}$ and $d^{\mu\nu}$ if they are considered together. In contrast, like proposed in \cite{Altschul:2006ts,Karki:2022hzm}, the results for the fields $f^{\mu}$ and $c^{\mu\nu}$ are related and equivalent, since \eqref{LLL6} is valid. As such a study was not made for $b^{\mu}$ and $d^{\mu\nu}$, we cannot infer that they could represent the same physical effect, we can only say that for the Fermi-Dirac distribution they represent the same shape, being one positive and the other negative.

We conclude by commenting that this analysis represent only a small evidence that we can try to find an relation between the fields $b^{\mu}$ and $d^{\mu\nu}$. For the $f^{\mu}$ and $c^{\mu\nu}$, the relation is established and we only contribute to enforce this idea.

\section*{Acknowledgements}

This study was financed in part by The Coordena\c c\~ao de Aperfei\c coamento de Pessoal de N\'ivel Superior - Brasil (CAPES) - Finance Code 001.

\bibliography{library}
\bibliographystyle{utphys2}

\end{document}